\documentclass[a4paper,11pt]{article}
\pdfoutput=1 
\usepackage{jheppub} 
\usepackage[T1]{fontenc}
\usepackage[italian,english]{babel}
\usepackage{hyperref}
\hypersetup{
	colorlinks=true,
	linkcolor=[rgb]{0.0,0.42,0.89},
	filecolor=magenta,      
	urlcolor=[rgb]{0.0,0.42,0.89},
	citecolor=[rgb]{0.64,0.0,0.0},
}
\usepackage{ifpdf}
\usepackage{subfigure}
\usepackage{gensymb}
\usepackage{amssymb}
\usepackage{amsfonts}
\usepackage{epsf}
\usepackage{rotating}
\usepackage{graphicx}
\usepackage{amsmath}
\usepackage{fancyhdr}
\usepackage{lineno}
\usepackage{babel, blindtext}
\usepackage{graphics}
\usepackage{color}
\usepackage{physics}
\usepackage{multirow}
\usepackage{pstricks}
\usepackage{xcolor}
\usepackage{multirow}
\usepackage{hhline}
\usepackage{cancel}
\usepackage{framed}
\usepackage{mathtools}
\usepackage{diagbox}
\usepackage{float}
\usepackage[lmargin=1.7in,bmargin=0.2in,footskip=0.5in,total={6.9in,9.5in}]{geometry}

\newcommand{\nn}{\nonumber}
\newcommand{\lsim}{\mathrel{\mathop{\kern 0pt \rlap
  {\raise.2ex\hbox{$<$}}}
  \lower.9ex\hbox{\kern-.190em $\sim$}}}
\newcommand{\gsim}{\mathrel{\mathop{\kern 0pt \rlap
  {\raise.2ex\hbox{$>$}}}
  \lower.9ex\hbox{\kern-.190em $\sim$}}}

\newcommand{\be}{\begin{equation}}
\newcommand{\ee}{\end{equation}}
\newcommand{\bea}{\begin{eqnarray}}
\newcommand{\eea}{\end{eqnarray}}

\title{\boldmath Electroweak phase transition with radiative symmetry breaking in Type-II seesaw with inert doublet}

\author[a]{Shilpa Jangid}
\author[b]{Hiroshi Okada}
\affiliation[a]{
Asia Pacific Center for Theoretical Physics (APCTP) - Headquarters San 31,
Hyoja-dong, Nam-gu, Pohang 790-784, Korea}
\affiliation[b]{
	Department of Physics, Kyushu University,
	744 Motooka, Nishi-ku, Fukuoka 819-0395, Japan}
\emailAdd{shilpa.jangid@apctp.org, okada.hiroshi@phys.kyushu-u.ac.jp}

\preprint{}

\abstract{We consider the Type-II seesaw model extended with another Higgs doublet, which is odd under the $Z_2$ symmetry. We look for the possibility of triggering the electroweak symmetry breaking via radiative effects. The Higgs mass parameter changes sign from being positive at higher energy scales to negative at lower energy scales in the presence of the TeV scalar triplet. The Planck scale perturbativity is demanded and the electroweak phase transition is studied using two-loop $\beta$-functions. The maximum allowed values for the interaction quartic coupling of the second doublet field and the triplet field with the Higgs field are $\lambda_3=0.15$ and $\lambda_{\Phi_{1\Delta}}=0.50$, respectively. Considering these EW values, the first-order phase transition, i.e., $\phi_{+}(T_c)/T_c\sim 0.6$ is satisfied only for vanishing doublet and triplet bare mass parameters, $m_{\Phi_2}=0.0$ GeV and $m_{\Delta}=0.0$ GeV. The small non-zero induced vacuum expectation value for the scalar triplet also generates the neutrino mass, and the lightest stable neutral particle from the inert doublet satisfies the dark matter constraints for the chosen parameter space. The impact of the thermal corrections on the stability of the electroweak vacuum is also studied, and the current experimental values of the Higgs mass and the top mass lie in the stable region both at the zero temperature and the finite temperature.}


\begin{document}

\maketitle
\flushbottom

\section{Introduction} 
The ATLAS \cite{ATLAS:2012yve} and CMS \cite{CMS:2012qbp} experiments at the Large Hadron Collider (LHC) made the momentous finding of the Higgs boson predicted by the Standard Model (SM), which resulted in a triumphant moment for particle physicists. This significant finding demonstrated that SM is the most effective theory for describing all fundamental interactions in particle physics. The SM physics up to accessible energies was really validated by this discovery, together with the electroweak precision evidence. Despite this, the evidence of the existence of non-zero neutrino mass and the predictions for a cold dark matter (DM) candidate by PLANCK \cite{Planck:2013pxb} and WMAP \cite{Correia:2010jf, WMAP:2012nax} explained the shortcomings of the Standard Model (SM). The possibility of non-zero mass for neutrinos is explained by different seesaw mechanisms. Out of which, the light neutrino masses are generated by the existence of heavy Majorona masses in the extension of SM with right-handed neutrinos (RHNs), i.e., Type-I seesaw \cite{Minkowski:1977sc,Sawada:1979dis,Gell-Mann:1979vob,Mohapatra:1979ia}. The other possibility is that the extension to a heavy scalar triplet and the small induced vacuum expectation value (vev) will generate the small neutrino mass, i.e., Type-II seesaw \cite{Schechter:1980gr,Lazarides:1980nt,Mohapatra:1980yp,Wetterich:1981bx,Schechter:1981cv}. Secondly, the simplest way to get a DM candidate is SM extended to scalar multiplets with a discrete $Z_2$ symmetry for inert doublet models \cite{Deshpande:1977rw,Jangid:2020qgo,Jangid:2020dqh}. This $Z_2$ symmetry can also be a remnant symmetry of higher gauge groups, i.e.,  $SO(10)$ grand unified theories (GUTs) \cite{Kibble:1982ae,Mohapatra:1986su,Font:1989ai,Krauss:1988zc,Ibanez:1991hv,Ibanez:1991pr,Martin:1992mq,Mambrini:2016dca,Held:2022hnw}. These higher gauge groups provide interesting physics at higher energies; unification of SM gauge couplings is one of them. \\

The possibility of the Higgs boson being a part of larger multiplets in various extensions breaks some higher symmetries. For example, in supersymmetric scenarios, for the extension to a $U(1)_{B-L}$ symmetry, the breaking scale of $U(1)_{B-L}$ can be brought down to the supersymmetric scale by choosing parameters in an appropriate way \cite{Khalil:2016lgy,Burell:2011jhh}. This high-scale symmetry breaking demands all physical Higgs bosons, including the SM Higgs, have positive mass squared values at higher energies. There is a way to trigger electroweak symmetry breaking (EWSB) by radiative effects in supersymmetric scenarios \cite{Ibanez:1982fr,Inoue:1982pi,Ibanez:1982ee,Ellis:1982wr,Ellis:1983bp,Alvarez-Gaume:1983drc}. In such cases, the Higgs mass parameter, which is positive at higher energies, can turn negative at lower energies via radiative loop corrections and lead to EWSB. The idea of radiative electroweak symmetry breaking in various non-susy scalar extensions, i.e., Type-II seesaw with TeV scalar triplet and dark matter models (DM) in singlet and doublet extensions of the SM, has been studied in detail \cite{Babu:2016gpg}.\\

Another missing thing in the SM is that the electroweak phase transition (EWPT) from the symmetric phase to the broken phase is a smooth crossover \cite{Aoki:1999fi} for the Higgs boson mass with more than 80 GeV \cite{Kajantie:1996mn,Kajantie:1996qd,Csikor:1998eu}, which is not at all consistent with the measured Higgs boson mass of 125.5 GeV \cite{Kajantie:1995kf}. This is another motivation for the extension of the SM with additional scalar degrees of freedom, minimally with singlet \cite{Espinosa:1993bs,Bandyopadhyay:2021ipw}, doublet \cite{Blinov:2015vma} or triplet \cite{Bandyopadhyay:2021ipw} in the $SU(2)$ representation. The strongly first-order phase transition (FOPT) in Type-II seesaw has already been discussed in detail 
\cite{Zhou:2022mlz,Roy:2022gop}. As clear from the various SM extensions to the singlet and the triplet, the EWPT is favored by lighter masses \cite{Bandyopadhyay:2021ipw,Jangid:2023jya}. Hence, it is interesting to look for the strongly first-order EWPT in accordance with the radiative EWSB. One possibility is to introduce two different multiplets; one of them can be chosen to be of higher mass to provide the radiative symmetry breaking, while the masses for the second multiplet can be considered small enough to accommodate the strongly first-order phase transition. Either of the multiplets can be considered odd under the $Z_2$ symmetry in order to accommodate the cold DM candidate. Hence, we consider the extension of the SM with TeV scalar triplet, which also serves the purpose of generating the neutrino mass with a tiny non-zero induced vev, i.e., Type-II seesaw, and a second $SU(2)$ Higgs  doublet, which is considered to be odd under the $Z_2$ symmetry. The lightest stable neutral particle would be the cold DM candidate. The extension of SM with singlet is also a possibility, but doublet is chosen because of the interesting phenomenology with restricted regions of the mass range satisfying the DM constraints.\\

Lastly, the theoretical incompatibility of the SM with the stability of the EW vacuum at zero temperature \cite{Buttazzo:2013uya,Degrassi:2012ry} leads us to another motivation for extension to scalar degrees of freedom. The additional scalar degrees of freedom from the inert doublet and TeV scalar triplet gives positive contribution to the effective quartic coupling and enhances the stability of the EW vacuum. The electroweak vacuum stability from a cosmological perspective has already been studied in detail \cite{Espinosa:2007qp,Espinosa:2015qea,Kobakhidze:2013tn,Enqvist:2013kaa,Fairbairn:2014zia,Enqvist:2014bua,Kobakhidze:2014xda,Herranen:2014cua,Kamada:2014ufa,Shkerin:2015exa}. The Higgs field may be forced to tunnel down to the true vacuum because of the quantum fluctuations during inflation. This situation can happen even before the inflation ends, but this is not realised if the reheating temperature is sufficiently large after inflation. Hence, the Universe is characterized by sufficiently high value of the temperature after the end of inflation \cite{Espinosa:2015qea}. As a consequence, the thermal corrections effect the stability of the electroweak vacuum \cite{Isidori:2001bm, DelleRose:2015bpo,PhysRevD.44.3620} triggering the tunneling between the false vacuum to the true vacuum of the potential \cite{Coleman:1977py,Callan:1977pt,Anderson:1990aa,Espinosa:1995se} and cannot be neglected. Since, we are considering the finite-temperature corrections to the effective potential, it would be interesting to investigate the status for the stability of the EW vacuum at both zero temperature and the finite temperature. \\

The detailed outline of the article is as follows. The model description with EWSB and the mass expressions is given in \autoref{model}. The detailed description for the different possibilities of radiative EWSB is given in \autoref{rsb}. We check for the allowed parameter space from radiative symmetry breaking with strongly first-order phase transition in \autoref{ewpt} for the values allowed from the Planck scale perturbativity using two-loop $\beta-$ functions. The stability of the EW vacuum at zero temperature and the finite temperature is studied in \autoref{stability}. The detailed two-loop expressions for the quartic couplings and the gauge couplings are given in \autoref{betaf1}.

\section{Model setup}\label{model}
The minimal standard model (SM) is augmented with a $SU(2)$ Higgs triplet of $Y=1$, and a $SU(2)$ Higgs doublet of $Y=1/2$. The second doublet is considered to be odd under the $Z_2$ symmetry named "inert doublet (ID)", while the other fields are considered to be even under the $Z_2$ symmetry. The minimal Type-II Seesaw is considered to provide the radiative EWSB for the TeV mass scale of scalar triplet. The motivation for such a triplet comes from Left-Right symmetric theories, which can be perceived either at low-energy or can be realized in Grand Unified Theories (GUTs) or $E_6$. The non-zero vev of the Higgs triplet also provides the neutrino masses, and the lightest stable $Z_2$ odd particle from the inert doublet becomes the cold DM candidate.
The field definition for the Higgs multiplets is given as follows;
\bea
Z_2:  \Phi_1 \rightarrow \Phi_1, \Delta \rightarrow \Delta, \Phi_2 \rightarrow -\Phi_2,
\eea
where,
\begin{center}
	$	\Phi_1
	= \left(\begin{array}{c}
		G^+   \\
		\frac{1}{\sqrt{2}}(v_h+\rho_1+i G^0)  \end{array}\right) $, \qquad \qquad
	$\Delta =\frac{1}{\sqrt{2}} \left(
	\begin{array}{cc}
		\delta^+ & \sqrt{2} \delta^{++} \\
		(v_\Delta + \delta^0 + i \eta^0) & -\delta^+ \\
	\end{array}
	\right)$, \qquad \qquad $	\Phi_2
	= \left(\begin{array}{c}
		\phi^+   \\
		\frac{1}{\sqrt{2}}(H^0 + i A^0)  \end{array}\right).$
\end{center}
where $\Phi_2$ being odd under the $Z_2$ symmetry does not take part in the EWSB, and the mass eigenstates will be the same as the gauge eigenstates.
The scalar potential for the minimal Type-II seesaw scenario with additional inert doublet is as follows;
\begin{align}\label{eq:2.2}
V( \Phi_1, \Delta, \Phi_2) \ = \ & m_{\Phi_1}^2 (\Phi_1^{\dagger}\Phi_1)+ \lambda_{\Phi1} (\Phi_1^{\dagger}\Phi_1)^2 + m_{\Phi_2}^2 (\Phi_2^{\dagger}\Phi_2) + \lambda_{2}(\Phi_2^{\dagger}\Phi_2)^2+ m_{\Delta}^2 Tr(\Delta^{\dagger}\Delta) + [\mu_1(\Phi_1^T i \sigma_2 \Delta^{\dagger}\Phi_1)+h.c.]  \nn \\ & + \mu_2[\Phi_2^T i \sigma_2 \Delta^{\dagger}\Phi_2] +\lambda_{\Delta1}(Tr(\Delta^{\dagger}\Delta))^2   + \lambda_{\Delta2}Tr(\Delta^{\dagger}\Delta)^2 
 +\lambda_{\Phi_1 \Delta}(\Phi_1^{\dagger}\Phi_1)Tr(\Delta^{\dagger}\Delta)+ \lambda_{\Phi_1 \Delta1} \Phi_1^{\dagger}\Delta \Delta^{\dagger}\Phi_1 \nn \\ & + \lambda_3 \Phi_1^{\dagger}\Phi_1 \Phi_2^{\dagger}\Phi_2 + \lambda_4 \Phi_1^{\dagger}\Phi_2 \Phi_2^{\dagger}\Phi_1  + \frac{\lambda_5}{2}[(\Phi_1^{\dagger}\Phi_2)^2 + h.c.] +\lambda_{\Phi_2\Delta}(\Phi_2^{\dagger}\Phi_2)Tr(\Delta^{\dagger}\Delta) + \lambda_{\Phi_2 \Delta1} (\Phi_2^{\dagger}\Delta  \Delta^{\dagger}\Phi_2),
\end{align}
where, $Tr$ is the trace over the $2 \times 2$ matrix, and $\sigma_2$ is the second Pauli spin matrix. The triplet vev is restricted by the $\rho$ parameter $\rho = \frac{v_h^2 + 2 v_{\Delta}^2}{v_h^2 + 4 v_{\Delta}^2} \simeq 1-\frac{2 v_{\Delta}^2}{v_h^2}$ where the experimental measured value $\rho=1.00038_{-0.00020}^{+0.00020}$ and $\sqrt{v_h^2+v^2_\Delta}\approx$246.0 GeV restrict the triplet vev $ 0 \leq v_{\Delta}/{\rm GeV} \lsim 2.56$  implying that $v_{\Delta} \ll v_h$. The $\lambda_{\Phi_{1}\Delta}$ and $\lambda_{\Phi_{1}\Delta1}$ terms allow for the mixing between the SM Higgs and the scalar triplet for non-zero vev for the triplet Higgs. Rewriting the scalar potential given in \autoref{eq:2.2} in terms of the vevs and using the minimization conditions, the masses for the new scalar particles after spontaneous symmetry breaking are computed as follows \cite{Arhrib:2011uy};
\bea
m_{\Phi_1}^2 & = & \frac{1}{2}(2 \lambda_{\Phi_1} v_h^2 + (\lambda_{\Phi_1\Delta}+\lambda_{\Phi_1\Delta1})v_\Delta^2 -2 \sqrt{2}\mu_1 v_\Delta),\\
m_{\Delta}^2 & = & \frac{-(\lambda_{\Phi_1\Delta}+\lambda_{\Phi_1\Delta1})v_h^2 v_{\Delta}-2(\lambda_{\Delta1}+\lambda_{\Delta2})v_{\Delta}^3 + \sqrt{2} \mu_1 v_h^2}{2 v_{\Delta}}.
\eea
The doubly charged Higgs mass can be computed directly by collecting the coefficients for $\delta^{++}\delta^{--}$ in the scalar potential, and the mass expression for the doubly charged boson comes out to be:
\bea
m^2_{H^{\pm\pm}} = \frac{\mu_1 v_h^2 }{\sqrt{2}v_{\Delta}} -\frac{1}{2}\lambda_{\Phi_1\Delta1} v_h^2 - \lambda_{\Delta2} v_{\Delta}^2.
\eea 
The mass-squared mixing matrix for the singly charged Higgs is given as:
\begin{center}
	$	\mathcal{M}_{\pm}^2 
	= (\sqrt{2}\mu_1 -\frac{\lambda_{\Phi_1\Delta1}v_\Delta}{2}) \left(\begin{array}{cc}
		v_{\Delta} & - v_h/\sqrt{2}   \\
	- v_h/\sqrt{2} & v_h^2/ 2 v_{\Delta} \end{array}\right) $.
\end{center}
This matrix can be diagonalized using an orthogonal matrix, and one of the eigenvalues will be zero, which corresponds to the charged Goldstone degrees of freedom, and the non-zero eigenvalue would be the mass for the charged Higgs and is given by:
\begin{center}
	$	\mathcal{R}_{\beta^{\pm}} 
	=  \left(\begin{array}{cc}
	\cos\beta_{\pm} & \sin\beta_{\pm}   \\
		-\sin\beta_{\pm} & \cos\beta_{\pm} \end{array}\right) $, \qquad \qquad 
	$m^2_{H^{\pm}} 
	=  \frac{(v_h^2 + 2 v_{\Delta}^2)(2 \sqrt{2}\mu_1 -\lambda_{\Phi_1\Delta1} v_{\Delta})}{4 v_{\Delta}} $.
\end{center}
 In the similar way, the mixing matrix for the neutral CP-odd and CP-even scalars are given as;
 \begin{center}
 	$	\mathcal{M}^2_{CP_{even}} 
 	=  \left(\begin{array}{cc}
 		A & B   \\
 	B & C \end{array}\right) $, \qquad \qquad 
 	$\mathcal{M}^2_{CP_{odd}} 
 	= \sqrt{2}\mu_1 \left(\begin{array}{cc}
 		2 v_{\Delta} & -v_h   \\
 		v_h & v_h^2/2v_{\Delta} \end{array}\right) $,
 \end{center}
where
\bea
A= 2{\lambda_{\Phi_1}}v_h^2, \quad B= v_h((\lambda_{\Phi_1\Delta}+\lambda_{\Phi_1\Delta1})v_{\Delta}-\sqrt{2}\mu_1), \quad C= \frac{\sqrt{2} \mu_1 v_h^2 + 4(\lambda_{\Delta1}+\lambda_{\Delta2})v_{\Delta}^3}{2 v_{\Delta}}.
\eea
These matrices can be diagonalized by another rotation matrix, and for CP-odd mixing matrix, one of the eigenvalues would be zero, corresponding to neutral Goldstone degree of freedom, and the non-zero eigenvalue would be the mass for the neutral peudoscaler. The CP-even mixing matrix will give two physical eigenstates $h$ and $H$, where the lighter one $h$ is recognized as the SM Higgs boson, and the corresponding expressions are given below;

 \begin{center}
 	$	\mathcal{R}_{\beta^{0}} 
 	=  \left(\begin{array}{cc}
 		\cos\beta_{0} & \sin\beta_{0}   \\
 		-\sin\beta_{0} & \cos\beta_{0} \end{array}\right) $, \qquad \qquad $	\mathcal{R}_{\alpha} 
 	=  \left(\begin{array}{cc}
 	\cos\alpha_0 & \sin\alpha_0   \\
 	-\sin\alpha_0 & \cos\alpha_0 \end{array}\right) $,
 	
 \end{center}
 
 \bea
 m^2_{A} 
  & = &  \frac{\mu_1(v_h^2 + 4 v_{\Delta}^2)}{\sqrt{2}v_{\Delta}} ,\\
 m^2_h &  = & \frac{1}{2}[A+C -\sqrt{(A-C)^2 + 4 B^2}], \\
 m^2_H & = & \frac{1}{2}[A+C +\sqrt{(A-C)^2 + 4 B^2}].
\eea
Since the second doublet is inert and does not take part in the electroweak symmetry breaking, there would be no mixing, and the mass eigenstates would be the same as the gauge eigenstates. The physical masses for the second $SU(2)$ Higgs doublet would be in terms of the Higgs vev and the triplet vev, and the corresponding mass expressions are as follows:
\bea
m^2_{\phi^{\pm}} & = & m^2_{\Phi_2} + \frac{1}{2} \lambda_3 v_h^2 +\frac{1}{2}\lambda_{\Phi_2 \Delta}v_{\Delta}^2 \simeq m^2_{\Phi_2} + \frac{1}{2} \lambda_3 v_h^2, \\
m^2_{H^{0}} & = & m^2_{\Phi_2} + \frac{1}{2}(\lambda_3 +\lambda_4 + \lambda_5) +\frac{1}{2}(\lambda_{\Phi_2 \Delta} + \lambda_{\Phi_2 \Delta1})v_\Delta^2 - \sqrt{2}\mu_2 v_{\Delta} \simeq  m^2_{\Phi_2} + \frac{1}{2}(\lambda_3 +\lambda_4 + \lambda_5)v_h^2, \\
m^2_{A^{0}} & = & m^2_{\Phi_2} + \frac{1}{2}(\lambda_3 +\lambda_4 - \lambda_5) +\frac{1}{2}(\lambda_{\Phi_2 \Delta} + \lambda_{\Phi_2 \Delta1})v_\Delta^2 + \sqrt{2}\mu_2 v_{\Delta} \simeq  m^2_{\Phi_2} + \frac{1}{2}(\lambda_3 +\lambda_4 - \lambda_5)v_h^2.
\eea
The $SU(2)$ Higgs triplet generates Majorona mass terms for the neutrinos through the Yukawa interaction, as given below;
\bea\label{eq:2.5}
\mathcal{L}_Y = -\frac{(y_\nu)_{\alpha \beta}}{\sqrt{2}}l_i^T C  i \sigma_2 \Delta l_{\beta} + h.c.,
\eea
where, $\alpha$ and $\beta$ are the flavor indices for the leptons. The neutrino masses are generated by the small non-zero induced triplet vev, $v_{\Delta}= \frac{\mu_1 v_h^2}{\sqrt{2}m_{\Delta}^2} \ll v_h$ as given below \footnote{It is assumed that $m_{\Delta}^2 > 0$ implying that $v_{\Delta}$ is only induced after Higgs vev, $v_h$, is generated. In the case where $\mu_1$ is so small that $\mu_1 v_h^2  \ll m_{\Delta}^2$, the triplet scalar decouples. After integrating out the heavy scalar triplet, the Yukawa Lagrangian in \autoref{eq:2.5} generates an effective dimension-5 Weinberg operator to generate the neutrino mass.};
\bea
(m_\nu)_{\alpha \beta} \simeq \sqrt{2} v_{\Delta} (y_\nu)_{\alpha\beta}.
\eea
In the case of  Type-II seesaw, it is possible to acheive tiny neutrino masses even with $\mathcal{O}(1)$ neutrino Yukawa couplings because of the small vev for the Higgs triplet. This will be true simultaneously for light triplet scalars, even below TeV. In this article, we explore the possibility of radiative EWSB, which demands the additional scalar to be below a few TeV. The detailed analysis for the radiative EWSB is discussed in the next section.

\section{Radiative symmetry breaking}\label{rsb}
\begin{figure}[H]
	\begin{center}
		\mbox{\subfigure[BP1 ]{\includegraphics[width=0.5\linewidth,angle=-0]{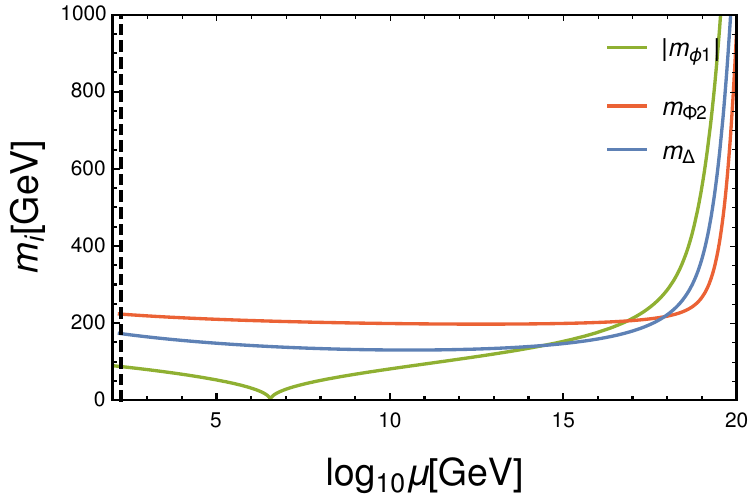}\label{f1a}}
			\subfigure[BP2]{\includegraphics[width=0.5\linewidth,angle=-0]{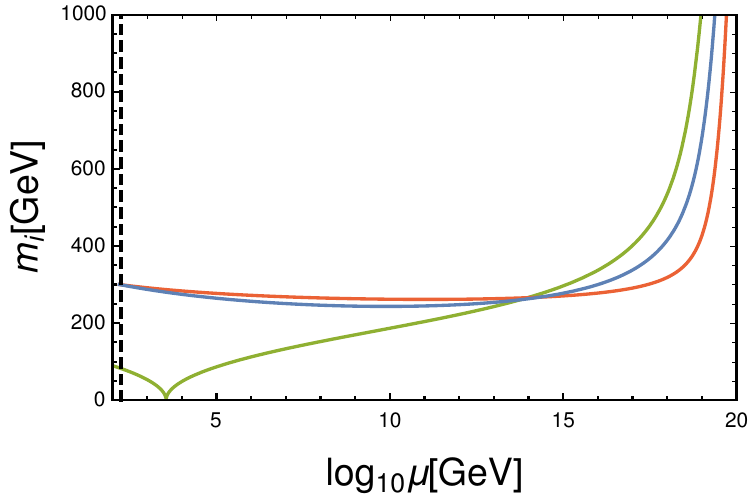}\label{f2a}}}
		\mbox{\subfigure[BP3 ]{\includegraphics[width=0.5\linewidth,angle=-0]{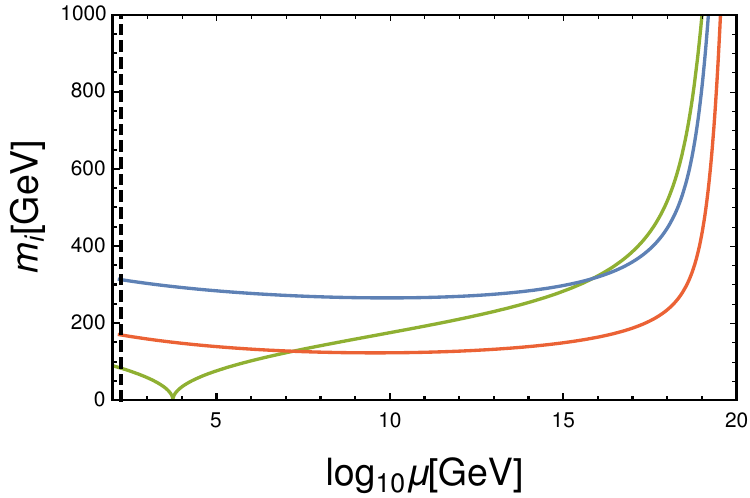}\label{f3a}}
			\label{f4a}}
		\caption{Variation of the bare mass parameter for the SM Higgs doublet, inert Higgs doublet and the $SU(2)$ triplet with the energy scale in GeV. The black dashed line corresponds to the scale equal to the mass of the lightest particle in the theory. The values of the different parameters needed in computing the radiative electroweak symmetry breaking are given in \autoref{table1} for three different benchmark points (BPs).}\label{fig:rsb}
	\end{center}
\end{figure}

The radiative EWSB is achieved when running of the Higgs mass parameter, $m_{\Phi_1}^2$  changes sign from positive to negative while evolving from high energy to low energies. The renormalization group flow for the Higgs mass parameter is proportional to itself as given in \autoref{SM} \cite{Arason:1991ic,Luo:2002ey}, which makes it impossible to turn negative at any energy scale. The magnitude of the Higgs mass parameter is $m_{\Phi_1}^2=-(88.0 \  \rm GeV)^2 $. In order to make this negative, the contribution of the additional scalar goes like $m_{multi}^2$ and is constrained to be not much heavier than TeV even if the interaction couplings are not very weak. 
\bea\label{SM}
\beta^{SM}_{m_{\Phi_1}^2}=\frac{1}{16 \pi^2}\Big[6 \lambda_{\Phi_1} - \frac{9}{10}g_1^2 -\frac{9}{2}g_2^2+ 2 Tr(3 Y_u^{\dagger}Y_u + 3 Y_d^{\dagger}Y_d + 3 Y_e^{\dagger}Y_e) \Big] m_{\Phi_1}^2.
\eea
The presence of a TeV scale triplet scalar in Type-II seesaw can aid in radiative symmetry breaking (RSB), and the masses for the second $SU(2)$ doublet can be assumed to be small. The expressions for the one-loop running of the mass parameters of this model are given as follows:
\bea
\beta_{m^2_{\Phi_1}} & = & \frac{1}{(16\pi)^2}\Big[\Big(-\frac{9}{10}g_1^2 -\frac{9}{2}g_2^2 + 12 \lambda_{\Phi_1} + 2T \Big)m^2_{\Phi_1} + \Big(4\lambda_3 + 2 \lambda_4\Big)m^2_{\Phi_2}   + \Big(6 \lambda_{\Phi_1\Delta} + 3 \lambda_{\Phi_1\Delta1}\Big)m_{\Delta}^2 +12 \mu_1^2\Big]; \label{eq:3.2}\\
\beta_{m^2_{\Phi_2}} & =  & \frac{1}{(16 \pi)^2}\Big[\Big(-\frac{9}{10}g_1^2 -\frac{9}{2}g_2^2 + 12 \lambda_{2} \Big) \Big)m^2_{\Phi_2}+ \Big(4\lambda_3 + 2 \lambda_4\Big)m^2_{\Phi_1}  + \Big(6 \lambda_{\Phi_2\Delta} + 3 \lambda_{\Phi_2\Delta1}\Big)m_{\Delta}^2+12 \mu_2^2\Big]; \\
\beta_{m^2_{\Delta}} & = & \frac{1}{(16\pi)^2}\Big[\Big(-\frac{18}{5}g_1^2 -12 g_2^2 + 16 \lambda_{\Delta1}+12 \lambda_{\Delta2}+4\mbox{Tr}\Big({Y_N  Y_{N}^{\dagger}}\Big) \Big)m_{\Delta}^2+\Big(4\lambda_{\Phi_1\Delta}+ 2\lambda_{\Phi_1\Delta1}\Big)m^2_{\Phi_1}\\
&&  +\Big(4\lambda_{\Phi_2\Delta}+ 2\lambda_{\Phi_2\Delta1}\Big)m^2_{\Phi_2} + 4\mu_1^2+4\mu_2^2\Big];\\
\beta_{\mu_1} & = & \frac{1}{(16\pi)^2} \Big[\Big(-\frac{27}{10}g_1^2 -\frac{21}{2}g_2^2 + 4 \lambda_{\Phi_1}+4 \lambda_{\Phi_1\Delta}+6 \lambda_{\Phi_1\Delta1}+ 2T  
+ 2 \mbox{Tr}\Big({Y_N  Y_{N}^{\dagger}}\Big)\Big)\mu_1 + 4 \lambda_5 \mu_2\Big];\\
\beta_{\mu_2} & = & \frac{1}{(16\pi)^2} \Big[\Big(-\frac{27}{10}g_1^2 -\frac{21}{2}g_2^2 + 4 \lambda_{2}+4 \lambda_{\Phi_2\Delta}+6 \lambda_{\Phi_2\Delta1}  
+ 2 \mbox{Tr}\Big({Y_N  Y_{N}^{\dagger}}\Big)\Big)\mu_2 + 4 \lambda_5 \mu_1\Big];
\eea
where
\bea
T= \rm Tr\Big[3 \mbox{Tr}\Big({Y_d  Y_{d}^{\dagger}}\Big)+  \mbox{Tr}\Big({Y_e  Y_{e}^{\dagger}}\Big) + 3 \mbox{Tr}\Big({Y_u  Y_{u}^{\dagger}}\Big)\Big].
\eea
The full two-loop $\beta$-functions for the quartic couplings are  computed using SARAH~\cite{Staub:2013tta} and are given in \autoref{betaf1}.
Following terms are crucial for changing the sign of Higgs mass parameter; $m_{\Phi1}$ can be $\Big(4\lambda_3 + 2 \lambda_4\Big)m^2_{\Phi_2} $ and $\Big(6 \lambda_{\Phi_1\Delta} + 3 \lambda_{\Phi_1\Delta1}\Big)m_{\Delta}^2$ or either it can be a combination of both from \autoref{eq:3.2}. Three different values for the mass parameters $m^2_{\Phi_2}$ and $m_{\Delta}^2$ are considered in \autoref{fig:rsb}, i.e, $m_{\Phi_2}=223.6,m_{\Delta}=173.2$ GeV,  $m_{\Phi_2}=300.0,m_{\Delta}=300.0$ GeV and $m_{\Phi_2}=173.3, m_{\Delta}=316.0$ in \autoref{fig:rsb}(a),(b) and (c), respectively.


The variation for the running mass parameters from Planck scale down to weak scale, i.e., $m_{\Phi1}, m_{\Phi2}$, and $m_{\Delta}$ is given in \autoref{fig:rsb}. The variation for the SM Higgs mass parameter, the additional $SU(2)$ Higgs doublet mass parameter, and the triplet bare mass parameter is denoted by the green, red, and the blue curves, respectively. The dotted black vertical line corresponds to the scale chosen equal to the mass of the lightest particle in the theory. The values of quartic couplings $\lambda_3, \lambda_4, \lambda_{\Phi_1 \Delta}$ and $\lambda_{\Phi_1\Delta1}$ are chosen to be equal for three different benchmark points to check the effect of different mass ranges for the bare mass parameters, and the corresponding values are also given in \autoref{table1}. These values of quartic couplings are the maximum allowed values at the EW scale for the Planck scale perturbativity. The Higgs mass parameter changes sign from positive to negative, i.e., ($m_{\Phi1}$ touches the horizontal line) at a particular energy scale depending on contributions from other bare mass parameters and the quartic couplings, i.e., $10^{6.5}$ GeV, $10^{3.6}$ GeV and $10^{3.8}$ GeV for BP1, BP2, and BP3 in \autoref{table1}, respectively. The radiative EWSB is triggered at these energy scales. The effect of triplet mass parameter is more crucial in triggering the radiative EWSB, since the value of quartic coupling $\lambda_{\Phi_1\Delta}$ is more compared to the interaction quartic coupling for the second doublet, i.e., $\lambda_3$. This higher value of the quartic coupling in multiplication with the bare mass parameter of the triplet enhances the positive contribution even more from the triplet. The scale for which the radiative symmetry breaking is triggered is reduced for more and more positive effect, as can be seen from \autoref{fig:rsb} (a)-(c). The mass parameter for the second doublet and the triplet are positive for all energy scales up to Planck scale, implying that the symmetry remains unbroken (since triplet vev is very small and induced only after the Higgs vev is generated). The mass parameter for the second doublet splits into $m_{\phi}^{\pm}, m_{H^0}$ and $m_{A^0}$, when the electroweak symmetry is broken.

\begin{table}[H]
	\begin{center}
		\renewcommand{\arraystretch}{1.4}
		\begin{tabular}{|c|c|c|c|c|c|c|}
			\hline
			\multirow{1}{*}{} & \multirow{1}{*}{BP1 }&
			\multicolumn{1}{|c|}{BP2}&	\multicolumn{1}{|c|}{BP3}\\
			\hline
			$m^2_{\Phi1}(\Lambda)/ ({\rm GeV})^2$ & $-(85.83)^2$ & $-(85.83)^2$ & $-(85.83)^2$\\
			\hline
			$m^2_{\Phi2}(\Lambda)/ ({\rm GeV})^2$  & $223.6^2$ & $300.0^2$ & $173.3^2$\\
			\hline
			$m^2_{\Delta}(\Lambda)/ ({\rm GeV})^2$ & $173.2^2$ & $300.0^2$ &  $316.0^2$ \\
			\hline
			$\lambda_3$  & 0.15 & 0.15 & 0.15\\
			\hline
			$\lambda_4$  & 0.00 & 0.00 & 0.00 \\
			\hline
			$\lambda_{\Phi_1 \Delta}$  & 0.50& 0.50 & 0.50\\
			\hline
			$\lambda_{\Phi_1 \Delta1}$ & 0.00& 0.00 & 0.00 \\
			\hline
		\end{tabular}
		\caption{The values relevant for the running of the bare mass parameters are provided for three benchmark points. $\Lambda$ denotes the scale corresponding to the mass of the lightest particle in the theory. The values of the quartic couplings chosen are allowed from the Planck scale perturbativity.}\label{table1}
	\end{center}
\end{table}

\begin{figure}[H]
	\begin{center}
		\mbox{\subfigure[ ]{\includegraphics[width=0.5\linewidth,angle=-0]{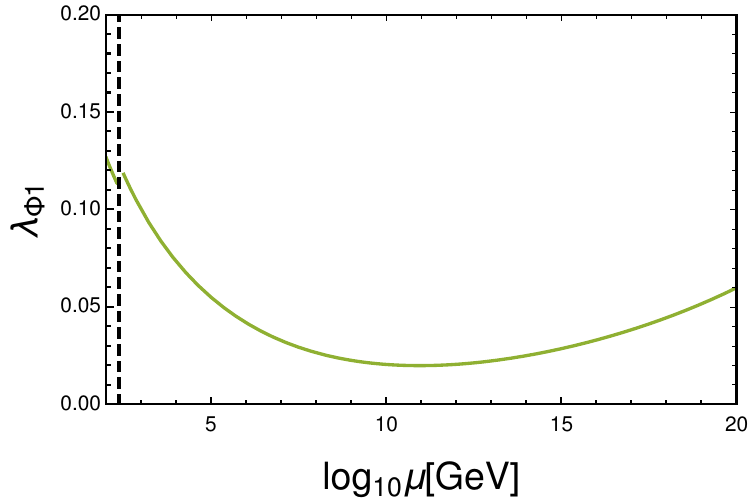}\label{f1a}}
			\subfigure[]{\includegraphics[width=0.5\linewidth,angle=-0]{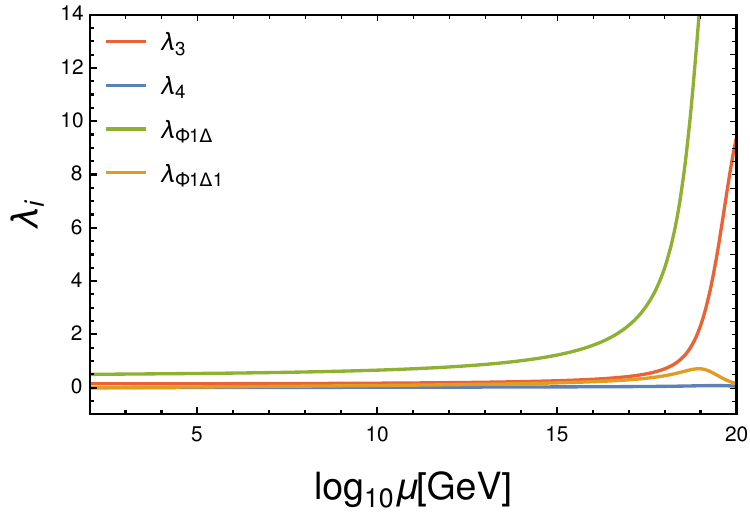}\label{f2a}}}
		\caption{\autoref{fig:quart}~(a) Variation of the interaction quartic couplings with the energy scale using two-loop $\beta$-functions. The Higgs quartic coupling variation is plotted in \autoref{fig:quart} (b), where the dashed black line corresponds to the scale equal to the mass of the lightest particle in the theory (similar for BP2 and BP3).}\label{fig:quart}
	\end{center}
\end{figure}

The variation of the quartic couplings relevant for the radiative EWSB and the Higgs quartic coupling using two-loop $\beta$-functions is given in \autoref{fig:quart}. The Planck scale perturbativity is achieved for all the quartic couplings (including those which are not plotted here). The running of the Higgs quartic coupling encounters a discontinuity at the scale $\Lambda$; above this energy scale, the full set of two-loop RGEs is used for the running of all the parameters of the theory. The values chosen at the EW scale  for the quartic couplings are same as given in \autoref{table1}.\\


The quartic couplings are chosen in terms of maximizing the value of the interaction coupling at the EW scale, which was possible more for the triplet case, not the inert doublet. This higher value of the interaction coupling would be very crucial in enhancing the phase transition dynamics. It would also be interesting to see how this interplay between the triplet mass and the inert doublet mass effects the strength of the phase transition. Hence, after discussing the radiative EWSB, the next section is devoted to EWPT using the RG-improved potential. 

\section{Electroweak phase transition(EWPT)}\label{ewpt}
The EWPT from the symmetric phase to the broken phase is accomplished by studying the finite-temperature effective potential in terms of background fields. The tree-level potential in \autoref{eq:2.2} includes contributions from one-loop Coleman Weinberg effective potential, i.e., $V_{\rm 1-loop}^{\rm CW}(\widetilde{m}_i^2)$, and one-loop potential at finite-temperature, i.e., $V_{\rm 1-loop}^{T \neq 0} (\widetilde{m}_i^2)$.  The full one-loop finite-temperature effective potential is given as follows:
\bea\label{eq:4.1}
V_1(T)= V_0^{\rm eff} + V_{\rm 1-loop}^{\rm CW}(\widetilde{m}_i^2) + V_{\rm 1-loop}^{T \neq 0} (\widetilde{m}_i^2),
\eea
where $V_0^{\rm eff}$ is the tree-level potential in \autoref{eq:2.2} rewritten in terms of background fields. 

 The one-loop effective potential at zero and finite temperature is given as follows \cite{Coleman};
\bea\label{eq:4.2}
V_{1-loop}^{\rm CW} & = & \frac{1}{(64\pi)^2}\sum_{i=B,F}(-1)^{F_i} n_i \hat{m_i}^4\Big[\log\Big(\frac{\widetilde{m}_i^2}{\mu^2}\Big)-k_i\Big], \\
V_{1-loop}^{T \neq 0} &  = & \frac{T^4}{(2\pi)^2}\sum_{i=B,F} (-1)^{F_i} n_i J_{B/F}\Big(\frac{\widetilde{m}_i^2}{T^2}\Big).
\eea
where, 
$\widetilde{m}_i^2 = \widetilde{m}_i^2(h_1,h_2,h_3; T) = \hat{m}_i^2(h_1,h_2,h_3) + \Pi_i T^2$
are the thermally corrected field dependent mass expressions. The field dependent mass expressions and the corresponding Debye corrections are given below:
\begin{center}
	$\mathcal{M}^2_{CP_{even}} 
	=  \left(\begin{array}{ccc}
		A_{11}& A_{12} & A_{13} \\
		A_{21} & A_{22} & A_{23} \\
		A_{31} & A_{32} & A_{33}\end{array}\right) $,
\end{center}
\bea
A_{11} & = & \frac{(\lambda_3+\lambda_4+\lambda_5)h_3^2}{2}+ 3 \lambda_{\Phi_1}h_1^2-\lambda_{\Phi_1}v_h^2+\frac{(\lambda_{\Phi_1 \Delta}+\lambda_{\Phi_1 \Delta1})h_2^2}{2}-\frac{(\lambda_{\Phi_1 \Delta}+\lambda_{\Phi_1 \Delta1})v_{\Delta}^2}{2} -\sqrt{2}\mu_1 h_2+\sqrt{2}\mu_1 v_{\Delta}, \nn \\
A_{12} & = & A_{21} =  (\lambda_{\Phi_1 \Delta}+\lambda_{\Phi_1 \Delta1}) h_1 h_2 -\sqrt{2}\mu_1 h_1, \qquad \qquad \qquad  A_{13}  =  A_{31} = (\lambda_3+\lambda_4+\lambda_5)h_1h_3, \nn \\
A_{22} & = & \frac{(\lambda_{\Phi_2 \Delta}+\lambda_{\Phi_2 \Delta1})h_3^2}{2}+\frac{(\lambda_{\Phi_1 \Delta}+\lambda_{\Phi_1 \Delta1})h_1^2}{2} - \frac{(\lambda_{\Phi_1 \Delta}+\lambda_{\Phi_1 \Delta1})v_h^2}{2} + 3(\lambda_{\Delta1}+\lambda_{\Delta2})h_2^2 -(\lambda_{\Delta1}+\lambda_{\Delta2})v_\Delta^2 +\frac{\mu_1 v_h^2}{\sqrt{2}v_{\Delta}}, \nn \\
A_{33} & = &  m_{\Phi_2}^2 +3 \lambda_2 h_3^2 + \frac{(\lambda_3+\lambda_4+\lambda_5)h_1^2}{2}  + \frac{(\lambda_{\Phi_2 \Delta}+\lambda_{\Phi_2 \Delta1})h_2^2}{2} -\sqrt{2}\mu_2 h_2, \nn \\
A_{23} & =& A_{32} = (\lambda_{\Phi_2 \Delta}+\lambda_{\Phi_2 \Delta1})h_2 h_3 - \sqrt{2}\mu_2 h_3.
\eea
In the similar way, the CP-odd mixing matrix in terms of the background fields is computed as;
\begin{center}
	$\mathcal{M}^2_{CP_{odd}} 
	=  \left(\begin{array}{ccc}
		B_{11}& B_{12} & B_{13} \\
		B_{21} & B_{22} & B_{23} \\
		B_{31} & B_{32} & B_{33}\end{array}\right) $,
\end{center}
\bea
B_{11} & = & \frac{(\lambda_3+\lambda_4-\lambda_5)h_3^2}{2}+  \lambda_{\Phi_1}h_1^2-\lambda_{\Phi_1}v_h^2+\frac{(\lambda_{\Phi_1 \Delta}+\lambda_{\Phi_1 \Delta1})h_2^2}{2}-\frac{(\lambda_{\Phi_1 \Delta}+\lambda_{\Phi_1 \Delta1})v_{\Delta}^2}{2} + \sqrt{2}\mu_1 h_2+\sqrt{2}\mu_1 v_{\Delta}, \nn \\
B_{12} & = & B_{21} =  -\sqrt{2}\mu_1 h_1, \qquad \qquad   B_{13}  =  B_{31} = 2\lambda_5 h_1h_3, \qquad \qquad   B_{23}  =  B_{32} = -\sqrt{2}\mu_2 h_3 , \nn \\
B_{22} & = & \frac{(\lambda_{\Phi_2 \Delta}+\lambda_{\Phi_2 \Delta1})h_3^2}{2}+\frac{(\lambda_{\Phi_1 \Delta}+\lambda_{\Phi_1 \Delta1})h_1^2}{2} - \frac{(\lambda_{\Phi_1 \Delta}+\lambda_{\Phi_1 \Delta1})v_h^2}{2} + (\lambda_{\Delta1}+\lambda_{\Delta2})h_2^2 -(\lambda_{\Delta1}+\lambda_{\Delta2})v_\Delta^2 +\frac{\mu_1 v_h^2}{\sqrt{2}v_{\Delta}}, \nn \\
B_{33} & = & m_{\Phi_2}^2 + \lambda_2 h_3^2 + \frac{(\lambda_3+\lambda_4-\lambda_5)h_1^2}{2}  + \frac{(\lambda_{\Phi_2 \Delta}+\lambda_{\Phi_2 \Delta1})h_2^2}{2} +\sqrt{2}\mu_2 h_2.
\eea
The field dependent charged mixing matrix is computed as follows;
\begin{center}
	$\mathcal{M}^2_{\pm} 
	=  \left(\begin{array}{ccc}
		C_{11}& C_{12} & C_{13} \\
		C_{21} & C_{22} & C_{23} \\
		C_{31} & C_{32} & C_{33}\end{array}\right) $,
\end{center}
\bea
C_{11} & = & \frac{\lambda_3h_3^2}{2}+  \lambda_{\Phi_1}h_1^2-\lambda_{\Phi_1}v_h^2+\frac{\lambda_{\Phi_1 \Delta}h_2^2}{2}-\frac{(\lambda_{\Phi_1 \Delta}+\lambda_{\Phi_1 \Delta1})v_{\Delta}^2}{2} +\sqrt{2}\mu_1 v_{\Delta}, \nn \\
C_{12} & = & C_{21} = \frac{\lambda_{\Phi_1 \Delta1}}{2}h_1 h_2 -\sqrt{2}\mu_1 h_1, \qquad \qquad   C_{13}  =  C_{31} = \frac{\lambda_4}{2}h_1 h_3 + \lambda_5 h_1h_3, \qquad \qquad   C_{23}  =  C_{32} = \frac{\lambda_{\Phi_2 \Delta1}}{2}h_2 h_3 -\sqrt{2}\mu_2 h_3 , \nn \\
C_{22} & = & \frac{(2\lambda_{\Phi_2 \Delta}+\lambda_{\Phi_2 \Delta1})h_3^2}{2}+\frac{(2\lambda_{\Phi_1 \Delta}+\lambda_{\Phi_1 \Delta1})h_1^2}{2} - (\lambda_{\Phi_1 \Delta}+\lambda_{\Phi_1 \Delta1})v_h^2 + (2\lambda_{\Delta1}+\lambda_{\Delta2})h_2^2 -2(\lambda_{\Delta1}+\lambda_{\Delta2})v_\Delta^2 +\frac{\sqrt{2}\mu_1 v_h^2}{v_{\Delta}}, \nn \\
C_{33} & = & m_{\Phi_2}^2 + \lambda_2 h_3^2 + \frac{\lambda_3 h_1^2}{2}  + \frac{\lambda_{\Phi_2 \Delta}h_2^2}{2} .
\eea
and
\bea
\hat{m}_{G^0}^2 = \lambda_h h_1^2 - m_h^2, \qquad \hat{m}_{W}^2 = \frac{g_2^2}{4} h_1^2, \qquad \hat{m}_{Z}^2 = \frac{g_2^2 + g_1^2}{4}h_1^2, \qquad \hat{m}_t^2 =\frac{y_t^2}{2}h_1^2, \nn
\eea
where, $\hat{m}_{G_0}, \hat{m}_{W}^2, \hat{m}_Z^2, \hat{m}_t^2$ are the masses for the Goldstone bosons, the gauge bosons, and the top quark, respectively. The $h_1,\ h_2$ and $h_3$ are the background fields for the SM Higgs doublet, additional triplet and the inert doublet, respectively.
The thermal corrections for the masses belonging to the same multiplet would be the same, and the corresponding expressions for the Debye coefficients are given below;
\bea\label{eq:4.7}
\Pi_{h} & = & \Big(\frac{g_1^2+3g_2^2}{16} + \frac{\lambda_{\Phi_1}}{2}+ \frac{y_t^2}{4}+ \frac{2\lambda_3+\lambda_4}{12} +\frac{2\lambda_{\Phi_1 \Delta}+\lambda_{\Phi_1 \Delta1}}{6} \Big)T^2, \nn \\
\Pi_{G^0} & = & \Big(\frac{g_1^2+3g_2^2}{16} + \frac{\lambda_{\Phi_1}}{2}+ \frac{y_t^2}{4}+ \frac{2\lambda_3+\lambda_4}{12} +\frac{2\lambda_{\Phi_1 \Delta}+\lambda_{\Phi_1 \Delta1}}{6} \Big)T^2, \nn \\
\Pi_{\Delta} & = & \Big(  \frac{2\lambda_{\Phi_1 \Delta}+\lambda_{\Phi_1 \Delta1}}{12} + \frac{2\lambda_{\Phi_2 \Delta}+\lambda_{\Phi_2 \Delta1}}{12} + \frac{(5\lambda_{\Delta1}+3\lambda_{\Delta2})}{6} \Big)T^2, \nn \\
\Pi_{\Phi_2} & = & \Big(  \frac{\lambda_{2}}{2} + \frac{2\lambda_3+\lambda_4}{12}+ + \frac{2\lambda_{\Phi_2 \Delta}+\lambda_{\Phi_2 \Delta1}}{6}  \Big)T^2, \nn \\
\Pi_{W_L} & = & \frac{7}{3}g_2^2 T^2, \nn \\
\Pi_{W_T} & = & \Pi_{Z_T}=\Pi_{\gamma_T} =0, \nn \\
\widetilde{m}_{Z_L}^2 & = & \frac{1}{2}\hat{m}_Z^2 +\frac{7}{6}(g_1^2+g_2^2)T^2 + \delta, \nn \\
\widetilde{m}_{\gamma_L}^2 & = & \frac{1}{2}\hat{m}_Z^2 +\frac{7}{6}(g_1^2+g_2^2)T^2 - \delta, \nn \\
\eea
where, $\Delta \in \{H,A,H^{\pm\pm},H^{\pm}\}$ and $\Phi_2 \in \{H^0,A^0,\phi^{\pm}\}$ as subscripts in \autoref{eq:4.7} consists of all the degrees of freedom for the additional triplet and the doublet. $\Pi_{h}, \ \Pi_{G^0}, \ \Pi_{\Delta}, \ \Pi_{\Phi_2}, \ \Pi_{W_L}, \ \Pi_{W_T} $ are the thermal corrections for the SM Higgs boson, goldstone boson, triplet, inert doublet, longitudinal and the transverse component of the $W$ boson, respectively. The thermal corrections for all the particles belonging to the same multiplet are same.  $\widetilde{m}_{Z_L}^2$ and $\widetilde{m}_{\gamma_L}^2$ are the Debye masses for the longitudinal and transverse component of $Z$ boson, and $\delta$ is computed as follows;
\bea
\delta^2= \Big(\frac{1}{2}\hat{m}_z^2+\frac{7}{6}(g_1^2+g_2^2)T^2\Big)^2 - \frac{7}{18}g_1^2 g_2^2 T^2(3 v_h^2 + 14 T^2).
\eea
After computing the masses in terms of background fields and the corresponding thermal corrections, the input parameters to be fed at the EW scale for two-loop $\beta$-functions are given in \autoref{tab:SMint} for computing the phase transition dynamics.

\begin{table}[H]
	\centering
	\begin{tabular}{|c|c|c|c|c|}
		\hline
		$g_1$&$g_2$& $g_3$&$y_t$&$\lambda_{\Phi_1}$\\
		\hline
		0.46256\footnotemark&0.64779&1.1666&0.93690&0.12604\\
		\hline
	\end{tabular}
	\caption{Initial values chosen at the EW scale for the RG evolution of the different SM parameters.}
	\label{tab:SMint}
\end{table}

\begin{figure}[H]
	\begin{center}
		\mbox{\subfigure[$Q=246$ GeV ]{\includegraphics[width=0.48\linewidth,angle=-0]{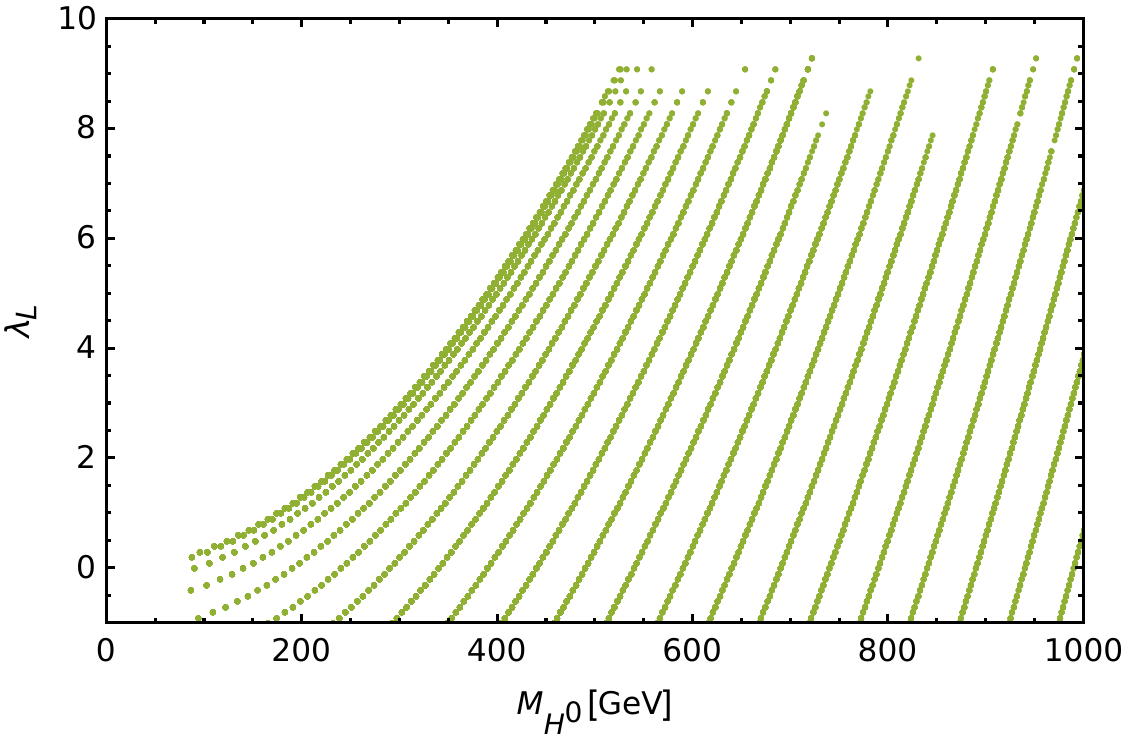}\label{f1a}}
			\subfigure[$Q=246$ GeV ]{\includegraphics[width=0.5\linewidth,angle=-0]{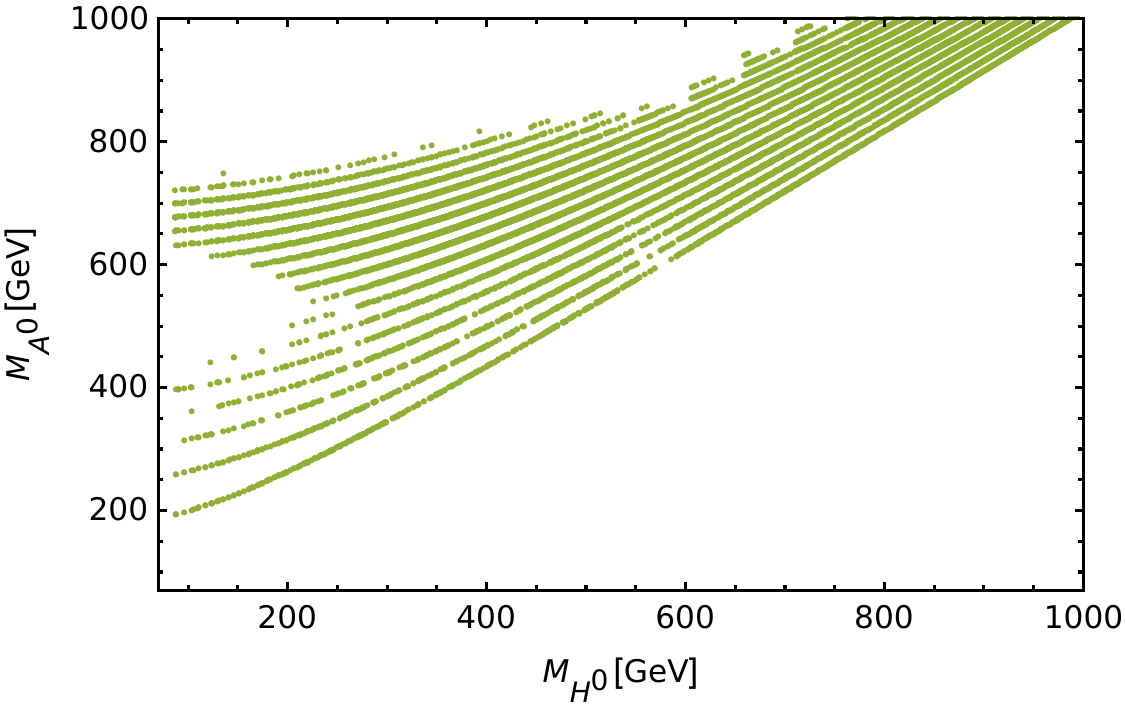}\label{f2a}}.}\caption{(a) Variation of the DM mass in GeV with the interaction coupling $\lambda_L=\lambda_3+\lambda_4+\lambda_5$, and (b) depicts the mass splitting between the DM mass $M_{H^0}$  and $M_{A^0}$ in GeV. The scattered points correspond to the strongly first-order phase transition and $Q=246$ GeV is the scale at which the running value of the couplings is considered for the strongly first-order phase transition. }\label{fig:246}
	\end{center}
\end{figure}

The one-loop Coleman-Weinberg potential has scale dependence in \autoref{eq:4.2}, and if the values chosen at the EW scale change a lot at higher scales, then it would be interesting to see the dynamics of the phase transition with the running couplings \cite{Blinov:2015vma}. The variation of the interaction coupling $\lambda_L=(\lambda_3+\lambda_4+\lambda_5)$ with the DM mass in GeV is shown in \autoref{fig:246} (a). The running of these parameters is considered until $\mu=Q=246$ GeV, and the corresponding values of these parameters at the scale $Q=246.0$ GeV are used as input for the EWPT.
The values of $\lambda_L$ are quite high but these values are perturbative till the considered energy scale of $Q=246$ GeV. Such high values of $\lambda_L$ will allow a strongly first-order phase transition for all possible values of the DM mass under consideration. The green points are all those points that strongly indicate a strongly  first-order phase transition, i.e., $\frac{\phi_+(T_c)}{T_c} \gsim 1.0$~\cite{Cohen:1993nk,Rubakov:1996vz} where $\frac{\phi_+(T_c)}{T_c}= \frac{\sqrt{h_1^2(T_c)+h_2^2(T_c)+(h_{3}(T_c)-h_{3}^h(T_c))^2}}{T_c}$ (in case of multiplets \cite{Ahriche:2014jna,Ahriche:2007jp}, when the false vacuum is $(0.0,0.0,h_{3}^h)$ instead of $(0.0,0.0,0.0)$. The mass splitting between the DM mass $M_{H^0}$ and $M_{A^0}$ is given in \autoref{fig:246} (b). The values of the other quartic couplings, $\lambda_i \in \{ \lambda_{\Phi_{1\Delta}}, \lambda_{\Phi_{1\Delta1}}, \lambda_{\Phi_{2\Delta}}, \lambda_{\Phi_{2\Delta1}}, \lambda_{\Delta1}, \lambda_{\Delta2}\}$ are chosen to be 0.01 and 0.1, and the solutions are the same for both the values.


The type-II seesaw extended with the inert doublet has a larger positive contribution to the running of couplings because of the more number of degrees of freedom. In that case, very low values for the quartic couplings would be allowed for achieving the Planck scale perturbativity, which would not be consistent with the strongly first-order phase transition. The same is also true for achieving perturbativity at any higher scale compared to $Q=246$ GeV. Hence, it would be interesting to see if the mass bounds triggering the radiative EWSB are also consistent with the strongly first-order phase transition and perturbative unitarity till that particular scale.\\

\begin{figure}[h]
	\begin{center}
			\mbox{\subfigure[Q =EW, $m_{\Phi_2}=0.0$ GeV ]{\includegraphics[width=0.48\linewidth,angle=-0]{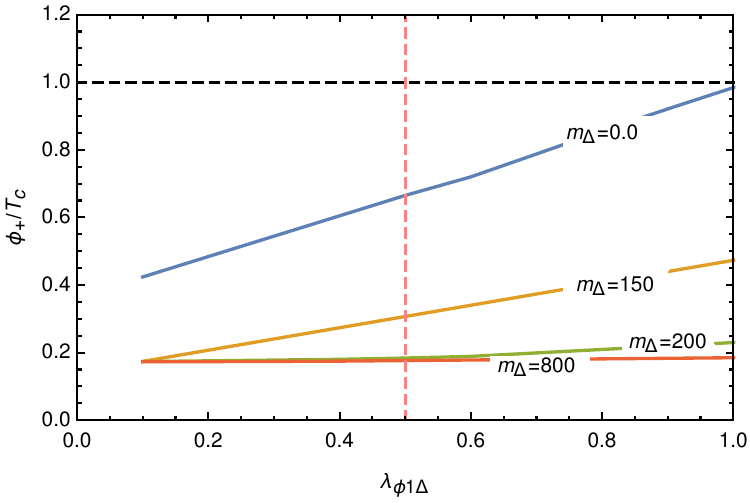}\label{f1a}}
			\subfigure[Q=EW, $m_{\Phi_2}=500.0$ GeV ]{\includegraphics[width=0.48\linewidth,angle=-0]{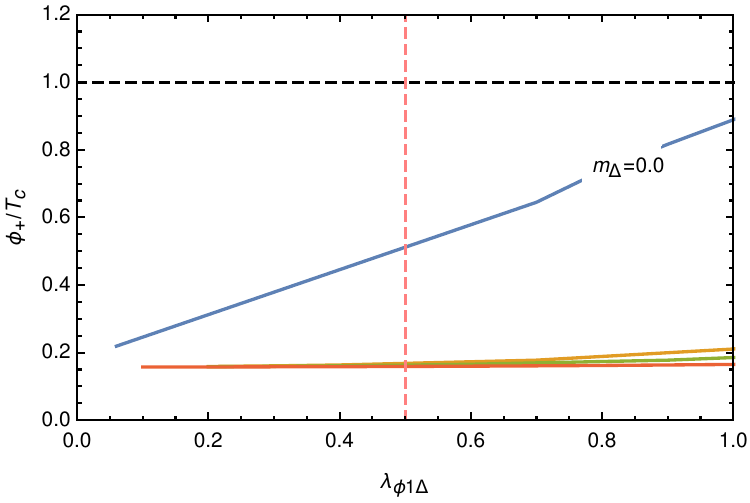}\label{f2a}}.}
		\caption{Variation of the strength of phase transition $\frac{\phi_+(T_c)}{T_c}$ with the interaction quartic coupling for SM Higgs and the triplet with different values of the mass parameter of the triplet. The inert doublet mass parameter is chosen as 0.0 GeV and 500.0 GeV in (a) and (b). (a) and (b) corresponds to the phase transition for the maximum allowed values of the quartic couplings at the EW scale that gives Planck scale perturbativity as given in \autoref{tab:int}.}\label{fig:1019}
	\end{center}
\end{figure}

For perturbative unitarity up to the Planck scale, we show the maximum allowed values for the quartic couplings at the EW scale, which are crucial for the strongly first-order phase transition are given in \autoref{tab:int}. \autoref{fig:1019} shows the variation of the interaction quartic coupling for the Higgs field and the triplet scalar field with the strength of phase transition $\frac{\phi_+(T_c)}{T_c}$. The black dashed horizontal line corresponds to the criteria for a strongly first-order phase transition. The dashed pink line corresponds to the maximum allowed value of the interaction quartic coupling at the EW scale in \autoref{fig:1019} (a)-(b). 
The strongly first-order phase transition is not achieved for any of the triplet masses, as shown in \autoref{fig:1019} (a) and (b) for two different values of the mass parameter of the inert doublet. The values at the EW scale are small enough and the strongly first order phase transition is not achieved for any of the doublet or the triplet masses. Still, the first-order phase transition is satisfied, i.e., $\frac{\phi_+(T_c)}{T_c} \gsim 0.6$ for the vanishing doublet and triplet bare mass parameters, $m_{\Phi_2}$ and $m_{\Delta}$ =0.0 GeV (\autoref{fig:1019} (a) and (b)).

\begin{table}[h]
	\centering
	\begin{tabular}{|c|c|c|c|c|c|c|c|c|c|c|c|}
		\hline
		 Q (GeV)&$\lambda_{\Phi_1}$&$\lambda_2$&  $\lambda_3$&$\lambda_4$&$\lambda_5$ & $\lambda_{\Phi_{1\Delta}}$ & $\lambda_{\Phi_{1\Delta1}}$ & $\lambda_{\Phi_{2\Delta}}$ & $\lambda_{\Phi_{2\Delta1}}$ & $\lambda_{\Delta1}$ & $\lambda_{\Delta2}$\\
		\hline
		EW &0.1264&0.01&0.15&0.00&-0.01 & 0.50 & 0.00 & 0.00 & -1.00 & 0.001 & 0.001\\
		\hline
	\end{tabular}
	\caption{The maximum allowed values of the quartic couplings from the Planck scale perturbativity at the EW scale.}
	\label{tab:int}
\end{table}

The allowed values of the quartic couplings at the EW scale which gives the Planck scale perturbativity will satisfy the first-order phase transition but this much lower masses cannot trigger the electroweak symmetry breaking radiatively.

\begin{table}[h]
	\centering
	\begin{tabular}{|c|c|c|c|c|c|c|c|c|c|c|c|}
		\hline
		Q (GeV)&$\lambda_{\Phi_1}$&$\lambda_2$&  $\lambda_3$&$\lambda_4$&$\lambda_5$ & $\lambda_{\Phi_{1\Delta}}$ & $\lambda_{\Phi_{1\Delta1}}$ & $\lambda_{\Phi_{2\Delta}}$ & $\lambda_{\Phi_{2\Delta1}}$ & $\lambda_{\Delta1}$ & $\lambda_{\Delta2}$\\
		\hline
		EW &0.1264&0.01&0.9&0.00&-0.01 & 0.90 & 0.00 & 0.00 & -1.00 & 0.001 & 0.001\\
		\hline
	\end{tabular}
	\caption{Benchmark points allowed for both radiative symmetry breaking and first-order phase transition.}
	\label{tab:int1}
\end{table}

\begin{figure}[H]
	\begin{center}
		\mbox{\subfigure[$m_{22}=100.0$ GeV ]{\includegraphics[width=0.48\linewidth,angle=-0]{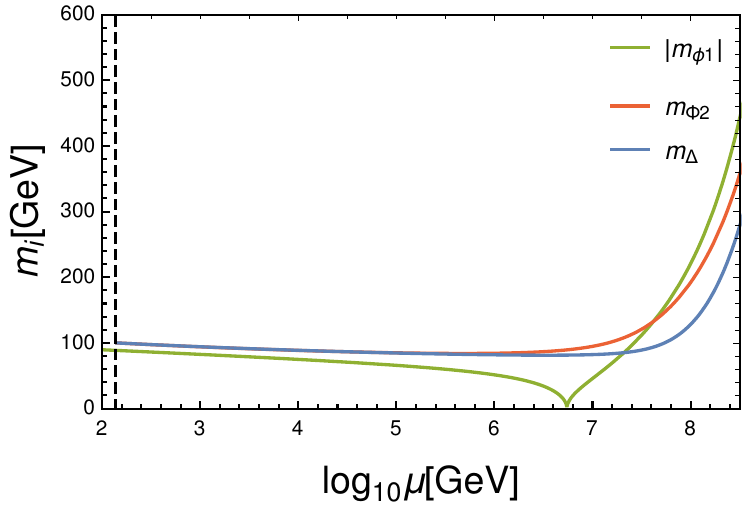}\label{f1a}}
			\subfigure[ $m_{22}=100.0$ GeV ]{\includegraphics[width=0.48\linewidth,angle=-0]{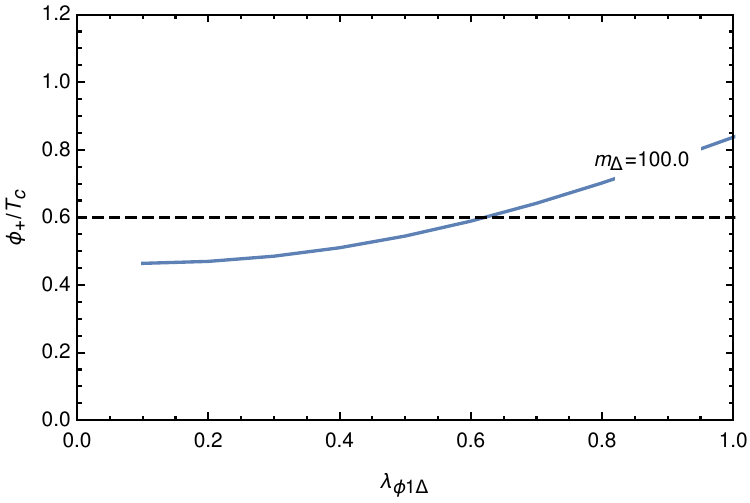}\label{f2a}}.}
		\caption{(a) Variation of the mass parameters with the scale for triplet and the inert doublet mass parameters both as 100.0 GeV. (b) Variation of the strength of phase transition $\frac{\phi_+(T_c)}{T_c}$ with the interaction quartic coupling for SM Higgs and the triplet. The triplet mass parameter and the doublet mass parameter are both chosen as $m_\Delta=100.0$ GeV and $m_{22}=100.0$ GeV, respectively. The black dashed line corresponds to the criteria for the first-order phase transition. }\label{fig:106}
	\end{center}
\end{figure}

\autoref{fig:106} (a) corresponds to the variation of the mass parameters of the theory with the energy scale. The values of the mass parameters are chosen to be $m_{\Delta}(\Lambda)$ and $m_{22}(\Lambda)$ = 100.0 GeV, and $\Lambda$ corresponds to the scale equal to the mass of the lightest particle in the theory as discussed in \autoref{rsb}. The electroweak symmetry breaking (EWSB) is triggered around $\sim 10^{6.80}$ GeV and the inert doublet mass parameter remains positive for all higher scales indicating that the $Z_2$ symmetry remains intact. The values of the relevant quartic couplings are chosen to be $\lambda_{3}=0.9$ and ${\lambda_{\Phi_{1\Delta}}=0.9}$ as given in \autoref{tab:int1},  which also satisfies the stability conditions since the interaction coupling $\lambda_3$ and $\lambda_{\phi1\Delta}$ are quite large as given in \cite{Bonilla:2015eha,Jangid:2020qgo,Jangid:2020dqh}. The same values are used for the variation of interaction quartic coupling $\lambda_{\Phi_{1\Delta}}$ with the strength of phase transition in \autoref{fig:106} (b). These are the minimum possible values for the mass parameters and the quartic couplings which satisfy both the first-order phase transition and radiative electroweak symmetry breaking. These bounds also satisfies the constraints from the direct detection in the DM region from 70 GeV to 1 TeV suggests $\mathcal{O}(0.01) \lsim |\lambda_L| \lsim \mathcal{O}(0.1)$ \cite{Kanemura:2010sh}. 

After discussing the EWPT, it is important to check for the stability of the EW vacuum by considering the running of all couplings till Planck scale. Since the phase transition dynamics includes thermal corrections, the thermal corrections also effect the tunneling probability from the false vacuum to the true vacuum. Hence, in the next section, we study stability of the EW vacuum both at the zero temperature, and the finite temperature.

\section{Finite temperature stability}\label{stability}
The electroweak vacuum stability till Planck scale is the theoretical drawback of the Standard Model within the top mass uncertainity. The tunneling probability from the electroweak minima to the second minima existing at the higher field values is triggered by the thermal corrections to the masses of the theory.
The decay lifetime of the electroweak vacuum is computed by considering the bounce solution to the classical equation of motion for the classical potential, $V(h_1)=\frac{\lambda_1}{4}h_1^4$ \cite{Kobakhidze:2013tn}. The zero temperature expression for the equation of motion is given as follows \cite{Coleman:1977py,Callan:1977pt};
\bea \label{eq:5.1}
\frac{d^2h_1}{dr^2} + \frac{3}{r}\frac{dh_1}{d r}=\frac{dV(h_1)}{dh_1}, \qquad  \lim_{r \rightarrow \infty}h_1(r)=0, \qquad \frac{dh_1}{dr}|_{r=0} =0,
\eea
with $r= |\vec{r}|$. The corresponding euclidean action for this $O(4)$ spherically symmetric solution is as follows;
\bea\label{eq:5.2}
S_E[h_1(r)]= 2\pi^2 \int_{0}^{\infty} dr r^3 \Big[\frac{1}{2}\Big(\frac{dh_1}{dr}\Big)^2 + V(h_1)\Big].
\eea
and the bounce solutions to the classical equation of motion in \autoref{eq:5.3} come out to be as:
\bea\label{eq:5.3}
h_{1B}(r)=\frac{8}{|\lambda_1|} \frac{R}{R^2+r^2},  \qquad \qquad \qquad S_E[h_{1b}(r)]=\frac{8\pi^2}{3 |\lambda_1|},
\eea
where $R$ the arbitrary scale characterizing the size of the bounce $(0 < R < \infty)$. This arbitrary parameter appears in \autoref{eq:5.3}, because of the approximation that the potential  considered above is scale invariant indicating that there is an infinite set of bounce solutions which lead to the same value for action. Addition of the quantum corrections break this scale invariance of the tree-level potential. This implies that the bounce solutions with different values of $R$ which used to give same action at the semi-classical level now gives a one-loop action as $S[h_{1b}(r)] \sim \frac{8 \pi^2}{(3 |\lambda(1/R)|)}$. Considering the effective potential as $V_{eff}=\frac{\lambda_{eff}}{4}h_1^4$, with $\lambda_{eff}$ is the effective quartic coupling including contributions from all the particles which are coupled to the Higgs field, the bounce solution in \autoref{eq:5.3} now becomes,
\bea\label{eq:5.4}
h_{1B}(r)=\frac{8}{|\lambda_{eff}(\mu)|} \frac{R}{R^2+r^2},  \qquad \qquad \qquad S_E[h_{1b}(r)]=\frac{8\pi^2}{3 |\lambda_{eff}(\mu)|}.
\eea
Now, there is just one particular value of $R$ which saturates the path integral, and defined as $R_M$. Considering the running of effective quartic coupling $\lambda_{eff}$, an instability is encountered when $\lambda_{eff}(\mu)$ hits zero, and then goes to negative values. The scale $\mu$ here is the renormalisation scale, ($R_M\sim 1/\mu$ is the size of the bounce) which minimizes the action, i.e., $\beta_{\lambda_{eff}}(\mu)=0$. After computing the scale $\mu$, it is easy to plot the profile of the bounce using \autoref{eq:5.4}. The field $h_1$ and the four-dimensional euclidean distance $r$ are both scaled using the Planck mass $M_P=1.22 \times 10^{19}$ GeV.
where, $\lambda_{eff}$ is the effective potential including contribution from all the particles  which are coupled to the Higgs field and is given by:
{\allowdisplaybreaks  \begin{align} \label{totalL}
		\lambda_{\rm eff}\left(h,\mu\right) & \ \simeq \  \underbrace{\lambda_h\left(\mu\right)}_{\text{tree-level}}+\underbrace{\frac{1}{16\pi^2}\sum_{\substack{i=W^\pm, Z, t, \\ h, G^\pm, G^0}} n_i\kappa_i^2 \Big[\log\frac{\kappa_i h^2}{\mu^2}-c_i\Big]}_{\text{Contribution from SM}}   
		+\underbrace{\frac{1}{16\pi^2}\sum_{\substack{i=\phi^\pm, \\ H^0, A^0}}n_i\kappa_i^2 \Big[\log\frac{\kappa_i h^2}{\mu^2}-c_i\Big]}_{\text{ Contribution from inert doublet  }}+\underbrace{\frac{1}{16\pi^2}\sum_{\substack{i=H^{\pm\pm}, A,\\ H, H^{\pm}}}n_i\kappa_i^2 \Big[\log\frac{\kappa_i h^2}{\mu^2}-c_i\Big]}_{\text{ Contribution from Type-II seesaw  }},
\end{align}}

Since, we have studied the electroweak phase transition from the symmetric phase to the broken phase using the finite temperature analysis, it would be interesting to see the effect of thermal corrections on the stability of the vacuum. In that case we use the full one-loop thermally corrected effective potential as given in \autoref{eq:4.1}.

The one-loop thermally corrected effective potential can be rewritten as the contribution of $V_{ 1-loop}^{T \neq 0}$ and $V_{ring}^{T\neq 0}$ as follows;
\bea
V_{\rm 1-loop}^{T\neq 0} = \sum_{i=W,G^0,Z,h,\Delta,\Phi_2} \frac{n_i T^4}{ 2\pi^2}J_B\Big(\frac{\hat{m}_i^2}{T^2}\Big) + \frac{n_tT^4}{2 \pi^2} J_F\Big(\frac{\hat{m}_t^2}{T^2}\Big), \label{eq:5.6}\\ 
V_{\rm ring}^{T\neq 0} = \sum_{i=W_L,G^0,Z_L,\gamma_L,h,\Delta,\Phi_2} \frac{n_iT^4}{12 \pi}\Big\{\Big[\frac{\hat{m}_i^2}{T^2}\Big]^{3/2}-\Big[\frac{\mathcal{M}_i^2}{T^2}\Big]^{3/2}\Big\} \label{eq:5.7},
\eea
where $\hat{m}_i^2$ are the masses in terms of the background filed and $n_i's$ are the degrees of freedom. One important thing to note is that the thermal corrections for the gauge bosons are non-zero only for the longitudinal degrees of freedom and there are no thermal corrections for the fermionic degrees of freedom. The expressions for the spline functions $J_{B,F}$ are defined as;
\bea
J_{B,F}(x^2) = \int_{0}^{\infty} dy y^2 \log(1 \mp e^{-\sqrt{y^2+x^2}}).
\eea
The zero temperature expression for the bounce given in \autoref{eq:5.1} is also modified at finite-temperature and is given as follows;
\bea \label{eq:6.0}
\frac{d^2h_1}{dr^2} + \frac{3}{r}\frac{dh_1}{d r}=\frac{dV_{eff}^{T\neq 0}(h_1)}{dh_1}, \qquad  \lim_{r \rightarrow \infty}h_1(r)=0, \qquad \frac{dh_1}{dr}|_{r=0} =0,
\eea
with $r= |\vec{r}|$. The corresponding euclidean action for this $O(3)$ spherically symmetric solution is as follows;
\bea
S_E[h_1(r)]= 4\pi \int_{0}^{\infty} dr r^2 \Big[\frac{1}{2}\Big(\frac{dh_1}{dr}\Big)^2 + V_{eff}^{T\neq 0}(h_1)\Big].
\eea
The effective potential can now be rewritten in high-temperature limit as follows;
\bea
V_{eff}^{T\neq 0} \simeq \frac{\lambda_{eff}}{4}h_1^4 + \frac{1}{2}\eta^2 h_1^2 T^2 + \rm const, \label{eq:5.11}
\eea
where the constant terms are neglected which are independent of the field $h_1$.
The coefficient for the field and the temperature dependence is given as;
\bea
\eta^2 & = & \frac{1}{12}\Big(\frac{3}{4}g_1^2 + \frac{9}{4} g_2^2 + 3 y_t^2 + 6 \lambda_{\Phi_1} + 2 \lambda_3 +\lambda_4 \Big)-\frac{\sqrt{2}}{32\pi}(g_1^3+3g_2^3) -  \frac{1}{16 \sqrt{3}\pi} (2\lambda_3 +\lambda_4)\sqrt{6\lambda_2 + (2\lambda_3 +\lambda_4) +  4\lambda_{\Phi_2 \Delta}+ 2 \lambda_{\Phi_2 \Delta1} } \nn \\
&& \quad - \frac{\sqrt{3}}{16 \pi}\lambda_{\Phi_1} \sqrt{3 g_1^2+9 g_  2^2 +24 \lambda_{\Phi_1}+12 y_t^2 + 8 \lambda_3 +4 \lambda_4 + 16\lambda_{\Phi_1 \Delta}+8\lambda_{\Phi_1 \Delta1} } \nn \\
&& \quad - \frac{\sqrt{3}}{16 \pi}\lambda_{\Phi_1 \Delta}\sqrt{2\lambda_{\Phi_1 \Delta}+\lambda_{\Phi_1 \Delta1}+ 2\lambda_{\Phi_2 \Delta}+\lambda_{\Phi_2 \Delta1} + (10\lambda_{\Delta1}+6\lambda_{\Delta2})} ,      
\eea
where the first term comes from the finite temperature one-loop potential in \autoref{eq:5.6} and the second and the third term comes from resummation in \autoref{eq:5.7}. Using the expression for effective potential in \autoref{eq:5.11}, the bounce solution at finite finite temperature can be straightway computed as $S_3[h_{1B}(r)] \simeq -(6.015)\pi \eta / \lambda_{eff}T$. It is important to note that all the couplings in the effective quartic coupling and $\eta$ are scale dependent and the running values are considered for all these parameters.

After computing the action, it is easy to evaluate the differential decay probability of the nucleating bubble at a particular temperature $T$ as follows;
\bea
\frac{dP}{d \ln T} \simeq \Gamma(T) \frac{M_P}{T^2}\Big(\frac{\tau_U T_0}{T}\Big)^3,
\eea
with $\Gamma(T)\simeq T^4 \Big\{ \frac{S_3[h_{1B}(r)]}{2 \pi T}\Big\}^{3/2} e^{-S_3[h_{1B}(r)]/T}$ is the vacuum decay rate per unit volume at a fixed temperature, $T$. $\tau_U$ is the age of the Universe and $T_0 \simeq 2.35 \times 10^{-4} $ eV. This differential decay probability is valid only in a radiation dominated Universe. 

The total integrated probability can be computed by integrating the differential decay probability as follows;
\bea
P(T_{\rm cut-off}) = \int_{0}^{T_{\rm cut-off}} \frac{dP(T')}{dT'}dT',
\eea
with $T_{\rm cut-off}$ is the maximum cut-off on the temperature which is decided by the validity till cut-off scale as $\Lambda$ when the bounce $h_{1B}(0)$ is computed at different temperatures.

\begin{figure}[H]
	\begin{center}
	 {\includegraphics[width=0.5\linewidth,angle=-0]{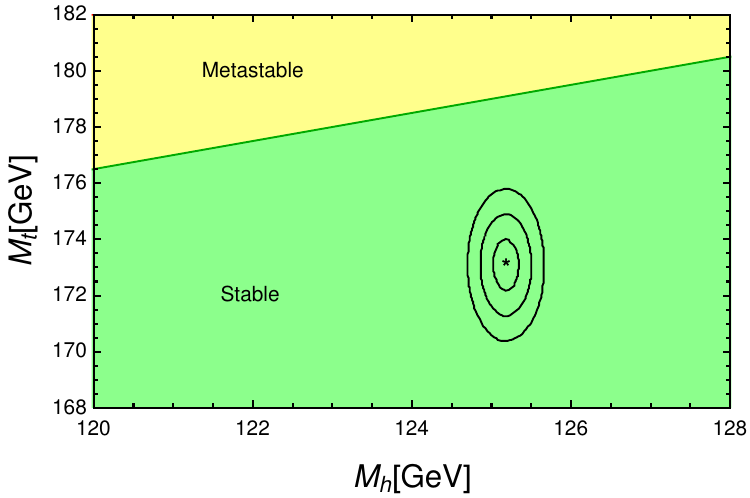}\label{f1a}}
		\caption{Phase space plot for the Higgs mass and the top mass in GeV at zero temperature. The green, yellow regions correspond to the stable and the metastable regions. The black contours denotes the $1 \sigma$, $2\sigma$ and $3 \sigma$ uncertainities and the black dot at the center corresponds to the current measured values of the Higgs mass and the top mass.}\label{fig:veffz}
	\end{center}
\end{figure}

The bounds on the tunneling probability can distinguish the regions of stability, metastability, and the instability. In case of second minima is higher than the EW minima, there is no possibility of tunneling, and this condition is defined as stable with criteria $\lambda_{eff}>0$. When the second minima lies below the EW minima, if the tunneling probability is greater than the age of the Universe, then the condition is defined as metastability. For the case, when the tunneling probability is less than the age of the Universe, then the region is unstable. The stable, metastable, and the unstable regions are denoted by green, yellow, and red colors in \autoref{fig:veffz}. The black contours correspond to $1 \sigma$, $2\sigma$ and $3 \sigma$ uncertainities, and the black dot at the center defines the experimentally measured values of the Higgs mass and the top mass. The current measured values of the Higgs mass and the top quark mass lie in the stable region because of the positive contribution from more number of scalar degrees of freedom both at the zero temperature and the finite temperature.

\section{Conclusion}
The extension of the Standard Model with a scalar triplet and a second $SU(2)$ Higgs doublet is considered with an additional $Z_2$ symmetry. The breaking of electroweak symmetry by radiative effects demands the mass of additional scalars to be below few TeV. But, the strongly first-order phase transition which is crucial for explaining the baryon asymmetry of the Universe, demands lower masses. The interaction quartic coupling of the additional multiplet with the Higgs field plays a very significant role in the strength of the phase transition. Higher is the value of this quartic coupling, higher would be the allowed mass from the strongly first-order phase transition. Hence, we have considered the extension of Type-II seesaw with another $SU(2)$ Higgs doublet where the masses of the inert doublet are chosen to be small enough to enhance the strength of the phase transition. In this way, it became possible to accommodate the additional doublet and the triplet scalar which provides the radiative electroweak symmetry breaking alongside the strongly first-order phase transition.\\

The Type-II Seesaw extended to inert doublet has more number of degrees of freedom which restricts the interaction coupling $\lambda_3$ for inert doublet to be $\sim 0.15$ and for the triplet scalar $\lambda_{\Phi_{1\Delta}}$ to be $\sim 0.5$ from Planck scale perturbativity. Hence, the effect on the scale where the electroweak symmetry breaking is triggered is more from the triplet in comparison to the doublet. Also, the maximum allowed values for the interaction couplings $0.15$ and $0.5$ for the doublet and the triplet are not enough to provide the strongly first-order phase transition. It can only provide the first-order phase transition $\phi_+(T_c)/T_c \gsim 0.6$ for doublet and triplet bare mass parameter to be $0.0$ GeV. Therefore, we have looked for the possibility for accommodating both radiative electroweak symmetry breaking and the first-order phase transition. The electroweak symmetry breaking is triggered around the scale $10^{6.80}$ GeV for  $m_{\Delta}(\Lambda)$ and $m_{22}(\Lambda)$ = 100.0 GeV and the first-order phase transition is also satisfied, i.e., $\frac{\phi_+(T_c)}{T_c} \gsim 0.6$.  \\

The stability of the electroweak vacuum is also studied both at the zero temperature and the finite temperature. The running of the couplings is considered until Planck scale for this analysis using two-loop $\beta$-functions. The positive contribution coming from the scalar degrees of freedom is too large and the measured values for the Higgs mass and the top mass lies in the stable region both at the zero temperature and the finite temperature.\\

The second doublet is considered to be odd under the $Z_2$ symmetry and the lightest stable neutral particle $M_{H^0}$ becomes the cold dark matter (DM) candidate. The DM relic density bound in case of inert doublet is satisfied in different mass ranges. Still, the DM constraints from DM relic in a low mass regime $M_{H^0} \lsim m_h/2$ and in resonant or funnel region, $M_{H^0} \sim m_h/2$ will also give the first-order phase transition and accommodate the radiative electroweak symmetry breaking. 

\section{Acknowlegement}
This research was supported by an appointment to the YST Program at the APCTP through the Science and Technology Promotion Fund and Lottery Fund of the Korean Government. This was also supported by the Korean Local Governments - Gyeongsangbuk-do Province and Pohang city (S.J.). S.J. thanks Luigi Delle Rose for useful guidance in computing the finite-temperature stability of the EW vacuum and Denis Comelli for computation of thermal corrections for the bosons.

\appendix
\section{Two-loop $\beta$-functions for dimensionless couplings} \label{betaf1}
\subsection{Scalar Quartic Couplings}\label{A1}
\footnotesize{
	\begingroup
	\allowdisplaybreaks

	\subsection{Gauge Couplings}
	\footnotesize{
		\begingroup
		\allowdisplaybreaks
		\begin{align*}
			\beta_{g_1} \  = \ &  
			\frac{1}{16\pi^2}\Bigg[\frac{24}{5} g_{1}^{3}\Bigg]+\frac{1}{(16\pi^2)^2}\Bigg[ \frac{1}{50} g_{1}^{3} \Big(-25 \mbox{Tr}\Big({Y_d  Y_{d}^{\dagger}}\Big)  + 424 g_{1}^{2}  + 440 g_{3}^{2}  -75 \mbox{Tr}\Big({Y_e  Y_{e}^{\dagger}}\Big)  -85 \mbox{Tr}\Big({Y_u  Y_{u}^{\dagger}}\Big)  + 900 g_{2}^{2}  -90 \mbox{Tr}\Big({Y_N  Y_N^\dagger}\Big) \Big)\Bigg] \, .  \\
			\beta_{g_2} \  = \ &  
			\frac{1}{16\pi^2}\Bigg[-\frac{7}{3} g_{2}^{3}\Bigg]+\frac{1}{(16\pi^2)^2}\Bigg[ \frac{1}{6} g_{2}^{3} \Big(160 g_{2}^{2}  -18 \mbox{Tr}\Big({Y_N  Y_N^\dagger}\Big)  + 36 g_{1}^{2}  -3 \mbox{Tr}\Big({Y_e  Y_{e}^{\dagger}}\Big)  + 72 g_{3}^{2}  -9 \mbox{Tr}\Big({Y_d  Y_{d}^{\dagger}}\Big)  -9 \mbox{Tr}\Big({Y_u  Y_{u}^{\dagger}}\Big) \Big)\Bigg] \, .  \\
			\beta_{g_3} \  = \ &  
			\frac{1}{16\pi^2}\Bigg[-7 g_{3}^{3}\Bigg]+\frac{1}{(16\pi^2)^2}\Bigg[-\frac{1}{10} g_{3}^{3} \Big(-11 g_{1}^{2}  + 20 \mbox{Tr}\Big({Y_d  Y_{d}^{\dagger}}\Big)  + 20 \mbox{Tr}\Big({Y_u  Y_{u}^{\dagger}}\Big)  + 260 g_{3}^{2}  -45 g_{2}^{2} \Big)\Bigg] \, .  \\
		\end{align*}
		\endgroup

		\bibliography{References}
		\bibliographystyle{Ref}
\end{document}